*Article*

# Low-cost sensors for indoor PV energy harvesting estimation based on machine learning

**Bastien Politi** [1,3] *****, **Alain Foucaran** [1] **and Nicolas Camara** [1, 2]

[1] Institute of Electronics and Systems, University of Montpellier, Montpellier, France;
[2] EPF – Graduate School of Engineering, Montpellier, France;
   nicolas.camara@epf.fr (N.C.);
[3] Bureaux A Partager SAS, Paris, France;
***** Correspondence: B.P. (Now at Hubert Curien Laboratory, University Jean Monnet, Saint-Etienne, France ;
   bastien.politi@univ-st-etienne.fr

**Abstract:** With the number of communicating sensors linked to the Internet of Things (IoT) ecosystem increasing dramatically, well-designed indoor light energy harvesting solutions are needed. The first step towards this development is to determine the harvestable energy in real indoor environments. But the harvestable energy varying over time with nature (spectra) and intensity of the light multi-sources, lighting data must be collected for sufficiently long periods. Besides, for a real implementation on-site, studies must be able to be carried out simultaneously in several places to determine locations with the highest energy harvesting potential. In this context, this manuscript presents a very low-cost prototype based on commercial photodiodes (rather than very expensive spectrometers), which measures only a very rudimentary number of spectral data. Thanks to a "classification" supervised machine learning from Matlab, in which an algorithm learns to classify new observations, and thanks to a simple principle of the superposition approximation model developed for flexible GaAs solar cells, our harvestable energy estimation error is less than 5% after more than 2 weeks of observation. To measure this error, the data collected leading to an estimate of the harvestable energy is compared to what has been experimentally harvested in a real IoT system Li-ion battery and compared to what has been estimated using an expensive spectrometer during the same period. Our prototype should allow the development and the massive deployment of a new generation of low cost 'indoor light energy harvesting' sensors for future reliable indoor energy harvesters.

## 1. Introduction

The development of the Internet of Things (IoT) or Wireless Sensor Networks (WSNs) applications is growing significantly. Reaching the energetic autonomy of the associated sensors remains a challenge. On one hand, a lot is done to reduce their power consumption, and on the other hand, a growing community of researchers works on improving the technologies able to harvest enough power from the nearby environment to supply electrical energy to such devices [1,2]. One of the most common means to achieve such energy recovery is photo-electrical conversion, based on photovoltaic technologies. For decades, databases and software have been used to accurately estimate the energy that can be harvested outdoors from square meters of sun light, wherever on earth [3]. On the contrary, under indoor light conditions, no standards have been created yet. Indeed, indoor light is usually composed of several different light sources (artificial and natural, direct and reflection). Therefore, characterization, standardization, and generalizations are more challenging in such environment than in outdoor light environments. That lack of standard makes the task of establishing the level of energy harvestable in those indoor conditions much more complex.

Nevertheless, many studies have been conducted to address this issue and some method capable of determining precisely the energy expected in indoor environments have been proposed [4,5]. These methods aim to help dimension an energy harvester accordingly to the needs of its application to result in an energetically autonomous device. But usually, only single light sources controlled in labs are considered for the models [6,7] and the studies mainly focusing on the PV performances at low light [1,8–13]. To go a step forward, it is now necessary to consider estimating the 'real' harvestable amount of electrical energy from a 'real' light in a 'real' indoor environment, and for a long period of time [14]. The 'real' harvestable amount of energy means considering the losses in non-ideal solar cell converters coupled to a non-ideal power management integrated circuit (PMIC) and non-ideal batteries. The efficiency of all these components that make the harvesting system must be evaluated to achieve a precise estimation. The 'real' light means to consider both artificial and natural light source including reflections and scatterings. Long periods must be investigated because of daily and seasonal light variations. Considering these elements have been already done in a study recently published by Politi et al. [15] where a spectrometer has been used for several weeks to quantify the harvestable energy in a specific location. The very good match between the estimation and the measurement of the energy harvested in the LiPo battery of a real electronic device after several days is promising for helping the engineering size their IoT energy harvesting systems. The main drawback of this technique is the price of a spectrometer and its needs for calibration and recalibration, which implies limitations in the number of days it can be deployed without maintenance and surveillance.

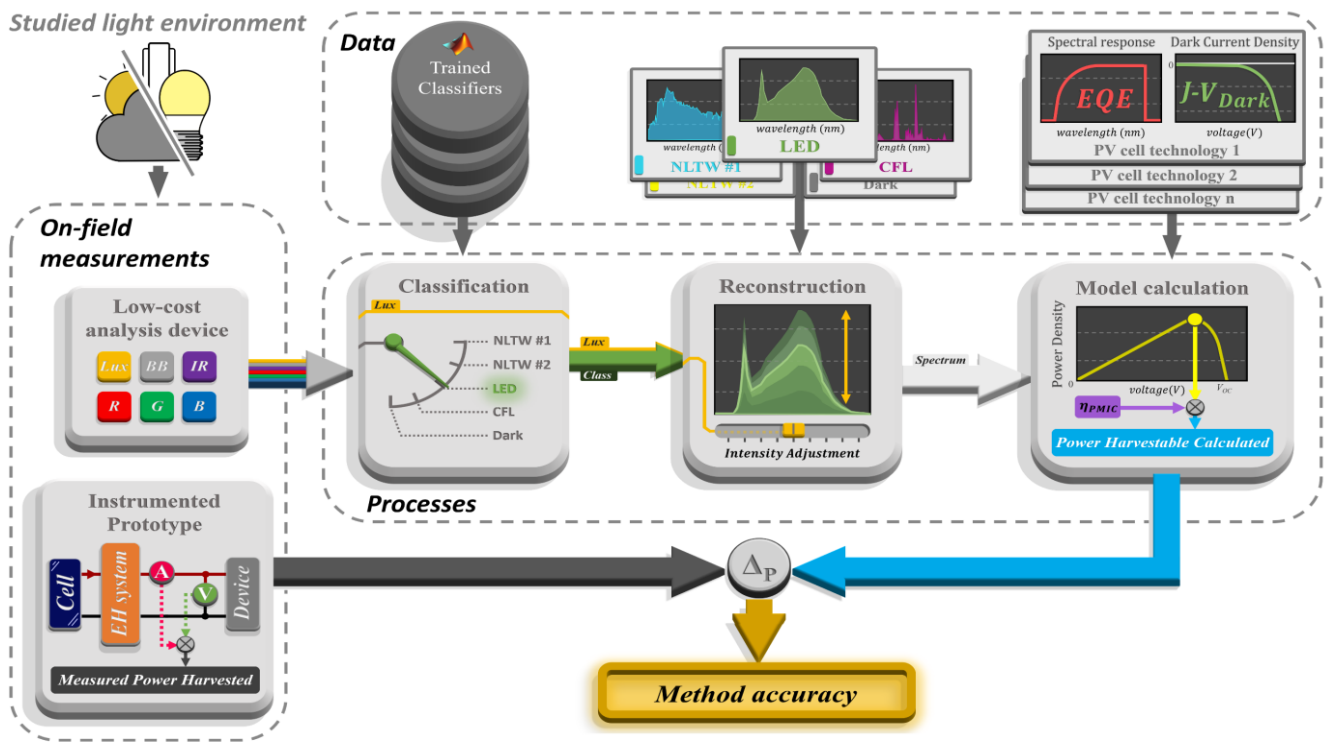

**Figure 1.** Descriptive diagram of the operational energy harvesting estimation method and its validation process. This method is based on multiple datasets created beforehand. The first one consists of successfully trained classifiers used to classify measurements made by the low-cost device (Classification Process) as one light class known (i.e., natural light through a window (NLTW), LED or CFL). The second one is the set of reference spectra taken while creating the training data for the classifiers. It is used to reconstruct a new spectrum from the class determined previously. The last dataset is composed of measurements made on the PV converter to be used. These data, associated with the reconstructed spectrum, can be fed to the calculation model for the final step of the method, the energy harvesting estimation.

In this methodological paper, we will present an innovative approach to achieve good performances on the study of the practical energy harvestable in real indoor environments, as reliable as in the report mentioned before [15], but at an ultra-low-cost price in the range of few dollars (compared to few thousands of dollars for a standard compact spectrometer). This system we have developed, based on the model and methods described in Politi et al. [15], inspired by the work of Sarik et al. [16] and Ma et al. [17], uses ultra-low-cost light sensor, associated with a machine-learning system to classify indoor light environments. Figure 1 shows the building blocks of our complete system which is detailed in the experimental setup section. In the following sections, we'll describe how we have selected the best classifier for our application and the energy harvesting estimation results we've obtained when choosing the best one, in real indoor light conditions.

## 2. Experimental Setup

### 2.1. Low-Cost Analysis Device

The main challenge regarding the method described in this article is to replace the expensive spectrometer described in Politi et al. by a low-cost low-tech system which can be widely deployed in the building to be tested. This system is composed of the combination of two sensors, widely available on a commercial basis and costing a few dollars each:

- The first sensor, the *TSL2561* from *TAOS* [19], is based on two photodiodes: i) the first one is a broadband (*BB*) photodiode, sensible to the whole visible + near infrared 300 nm to 1100 nm wavelengths, returning a digital *BB* value and ii) the second one is a narrower near infrared (*IR*) photodiode, sensible from 500 nm to 1100 nm wavelengths, returning a digital *IR* value. From these two photodiodes, an ambient light level value in lux can be derived using an empirical formula to approximate the human eye response. This value is returned as the digital *LUX* value.

- The second sensor, the *ISL29125* from *RENESAS* [20], uses a matrix of three photodiodes, each sensible to different parts of the visible spectrum: blue, green and red. It provides a set of three digital values *R*, *G*, and *B*.

A low-cost low-power *ESP32* microcontroller, able to communicate through WiFi, is used. Its purpose is to gather, treat and send the data via wireless communication protocols to a database which is stored in our server. This microcontroller is mounted on the commercially available *HUZZAH32* board (*Adafruit*) that integrates a USB connector used for battery charge and serial communication. An additional PCB has been designed, as a mother board to connect the two sensors to the ESP32 board, resulting in a compact device as small as 35 cm$^3$ shown in Figure 2(a-b).

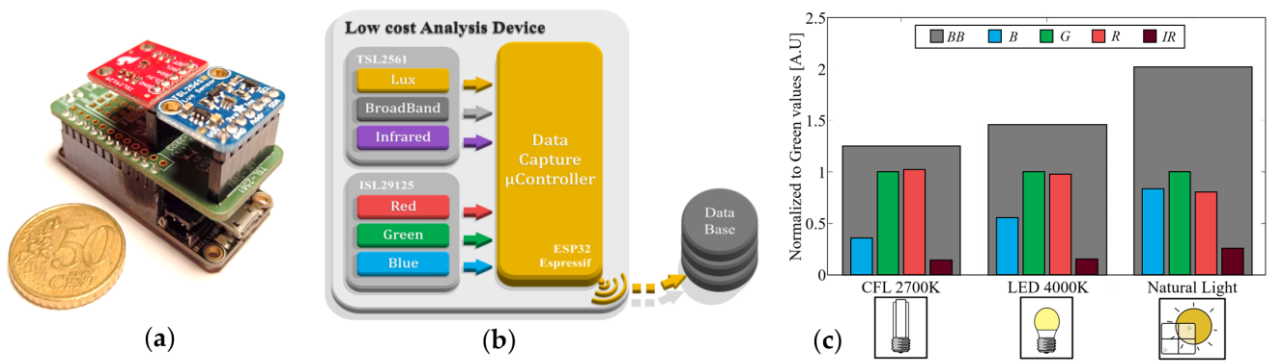

**Figure 2. (a)** Picture of the compact sensor (analysis system tool) that give light pieces of information via photodiodes dedicated to the broadband spectra, the infrared, the red, the green and the blue; **(b)** this information is collected and sent via an ESP32; **(c)** typical pseudospectra obtained under different light sources like CFL, LED or natural light through a window (NLTW).

This device can acquire what we call here "pseudo-spectra", which can be assimilated as very low-definition spectra. Even if the level of information provided by those pseudo-spectra is low, it remains sufficient to observe variations depending on the light source, as it can be seen on the example of Figure 1(c). If necessary pseudo-spectrum acquisition can be done (via USB or WiFi) at a relatively high rate (few measures per second).

*2.2. Classification and training*

From the low-resolution experimental pseudo-spectra provided by the low-cost device, machine learning algorithms are used to classify the different possible light sources. The method chosen for that purpose, called '*supervised machine learning classification*', is a classical Artificial Intelligence (AI) machine learning method. This method was implemented through the *Classification Learner App* made available by *MATLAB* in its *Statistics and Machine Learning* toolbox. In our study, we have divided the light sources into 7 classes: LED 3000 K, LED 4000 K, CFL 2700 K, CFL 6500 K, natural scattered light from a clean sky through a window, natural scattered light from a cloudy sky through a window and finally mere darkness. Varying the light intensity of each class, under controlled lab condition, we have built a dataset of 126 different pseudo spectra on which is based on our classification training. For each measurement of the different light source, we had to 'manually' inform the machine of the class of the tested light. Several classical classifiers included in the *Classification Learner app* can be simultaneously trained.

**Table 1.** List of the 24 algorithms tested as Classification method in this paper. These algorithms are the most classical one used in supervised machine learning for multinomial classification included in the *Classification Learner app* found in the *Statistics and Machine Learning* toolbox.

| Tree | Discriminant | Naive Bayes | Support Vector Machine (SVM) | K-nearest neighbor (KNN) | Ensembles |
|---|---|---|---|---|---|
| Fine | Linear | Gaussian | Linear | Fine | Boosted Trees |
| Medium | Quadratic | Kernal | Quadratic | Medium | Bagged Trees |
| Coarse | | | Cubic | Coarse | Subspace Discrimination |
| | | | Fine Gaussian | Cubic | Subspace KNN |
| | | | Medium Gaussian | Cosine | RUSBoosted Trees |
| | | | Coarse Gaussian | Weighted | |

Table 1 lists the 24 most classical algorithms used in supervised machine learning for multinomial classification. The main types of classification methods are the '*decision Trees*', the '*Discriminant analysis*', the '*Naïve Bayes*', the '*Support Vector Machine* (SVM)' Classification, the '*K-nearest neighbors* (KNN)' and finally a '*Classification Ensembles*'. These algorithms "learn" to classify data from training samples whose classes are known.

Thanks to the dataset of the 126 pseudo-spectra, each associated with one of the 7 defined classes, the training of the 24 classifiers can be done. It allows the building of the trained classifiers seen in the 'Data' part of the Figure 1. One of the challenges is to have classifiers reliable enough to be able to classify properly unknown new pseudo-spectra.

*2.3. Spectral Reconstruction and Calculation of the Harvestable Energy*

The spectrum database seen in the 'data' block of Figure 1 has been recorded thanks to a commercial calibrated spectroradiometer, the *StellarRAD* from *StellarNet Inc.*, equipped with a CR2 cosine receptor with a wavelength range from 350 nm to 1100 nm. The high resolution 7 spectra of the database, corresponding to the 7 classes used in our study, are necessary for the reconstruction stage one reliable trained classifier have given its classification results of new light environment observations. Then, depending on the class found by the proper classifier and the *LUX* value to tune the intensity properly, a high-resolution spectrum can be reconstructed.

Finally, to estimate the harvestable light energy in the surroundings of the low-cost sensor, we use a model, published in a previous article [15], that takes the reconstructed spectra as input instead of 'bare' high definition spectra from an expensive spectrometer. This model is based on a one-diode photovoltaic model using the measured external quantum efficiency (EQE) and the saturation current density–voltage characteristic measured in darkness (J-$V_{dark}$), both unchanging intrinsic values, depending only the PV converter technology to be tested. These EQE and J-$V_{dark}$ characteristics are contained in the database called '*Spectral response*' and '*Dark Current Density*' of the Data block of Figure 1. The spectral response was measured with a custom-built setup composed of a Xenon lamp, a monochromator equipped with two diffraction gratings, a filter wheel to remove the higher diffraction orders of radiation, and a lock-in amplifier. The measurements were calibrated with Si and Ge photodiodes (Thorlabs FDS100-CAL and FDG03-CAL, respectively) to cover the whole wavelength range of interest. The dark current density-voltage curves J-$V_{dark}$ of the different technologies PV converters have been collected by the SMU 2450 from Keithley.

*2.4. Instrumented prototype: Energy Harvester*

To evaluate the accuracy of the method presented below, it is mandatory to compare the energy harvesting calculation results with experimental measurements of the energy harvested by real energy harvesting systems. To perform such measurements an instrumented energy harvesting prototype has been created, based on 2 GaAs solar cells. It is embedded with power measurement units (*INA219* from *Texas Instrument Inc.*) capable of following the power harvested from solar cells by a power management integrated circuit (PMIC *AEM10941* produced by *ePEAS*) then stored into a battery. The picture shown in the Figure 3 displays this equipment and the data recovery system attached to it. In the same way as the low-cost light analysis tool, this system gathers data with a microcontroller which transfers its measurements to another microcontroller responsible for the data transmission to our server database. Then, accessing this database it is possible to analyze these data to confront them to the results of our low-cost method as seen in Figure 1.

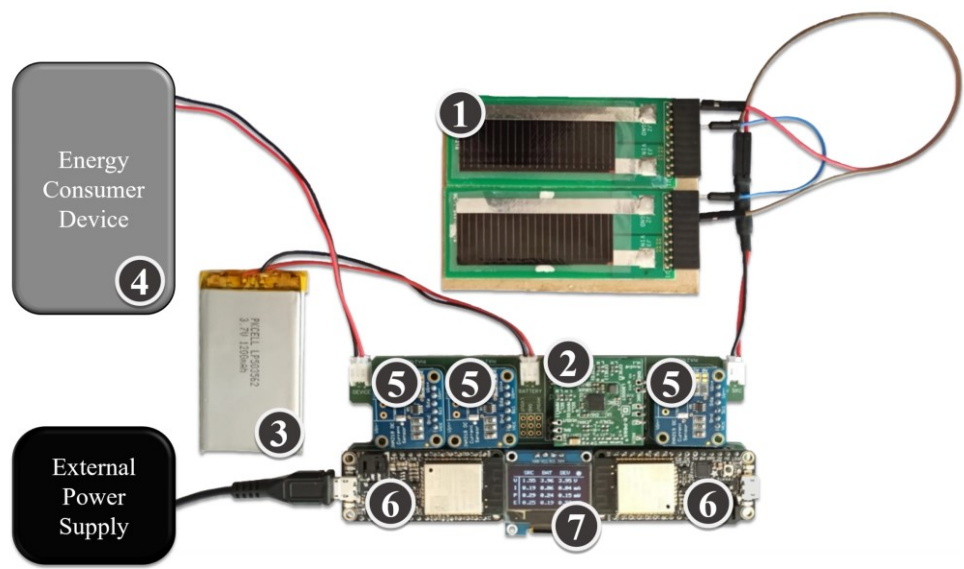

**Figure 3.** Picture of the experimental setup: (1) Two GaAs solar cells from *Alta Devices Inc.*; (2) Power Management Integrated Circuit (PMIC) from *e-PEAS*; (3) Lithium-Polymer battery of 4.4Wh; (4) Energy consuming devices; (5) INA219 power measurement integrated circuits from *Texas Instrument*; (6) ESP32 Microcontrollers in charge of gathering and storing data, integrated on HUZZAH Feather boards made by *Adafruit*; (7) OLED display to show system operating state.

## 3. Classification of Light Sources

*3.1. Classification Methods Training Performances on Raw Data*

Thanks to the *Statistics and Machine Learning* toolbox from Matlab and its *Classification Learner app*, classifiers can be trained very quickly. Those training can be performed with all six values of the pseudo-spectra but also with a reduced number of them. In our study, we have chosen the 11 configurations that we can see on the bottom of the Figure 4. For example, the configuration A relies on all the values from the sensors, while the configuration I utilizes only the Red (*R*) and InfraRed (*IR*) values from the sensors. Combining each of the 24 classification methods with each of these 11 configurations, it results in 264 classifiers to train. Those trainings begin with the 126 pseudo-spectra observed and each classifier uses a 5-fold cross validation. The cross-validation process consists of separating the dataset randomly into five folds. Four of them are then used to train the classifier. Once the training done, the classification capability of the classifier is tested on the remaining fold. This process is reproduced to obtain five iterations where each fold is used as the test fold. Finally, the overall performance of the classifier is find calculating the mean success rate of the five classification tests. Results of that process is shown in Figure 4 for the classifiers and configuration described previously.

A first observation is that none of the 264 classifiers reach a total 100% of classification success. A second look at the results shows that some classifiers seem more adapted to our application: Cubic SVM, Fine KNN or Weighted KNN. It is to be noted that the configurations with only 2 or 3 features of the dataset can perform almost as well as configurations with 5 or all data features. Finally, what can be concluded out of this example is that, with this dataset, the 'best' classifier is the Cubic SVM for the configuration A, which obtained 96.8% of correct classification: it was able to correctly classify 122 observations over 126.

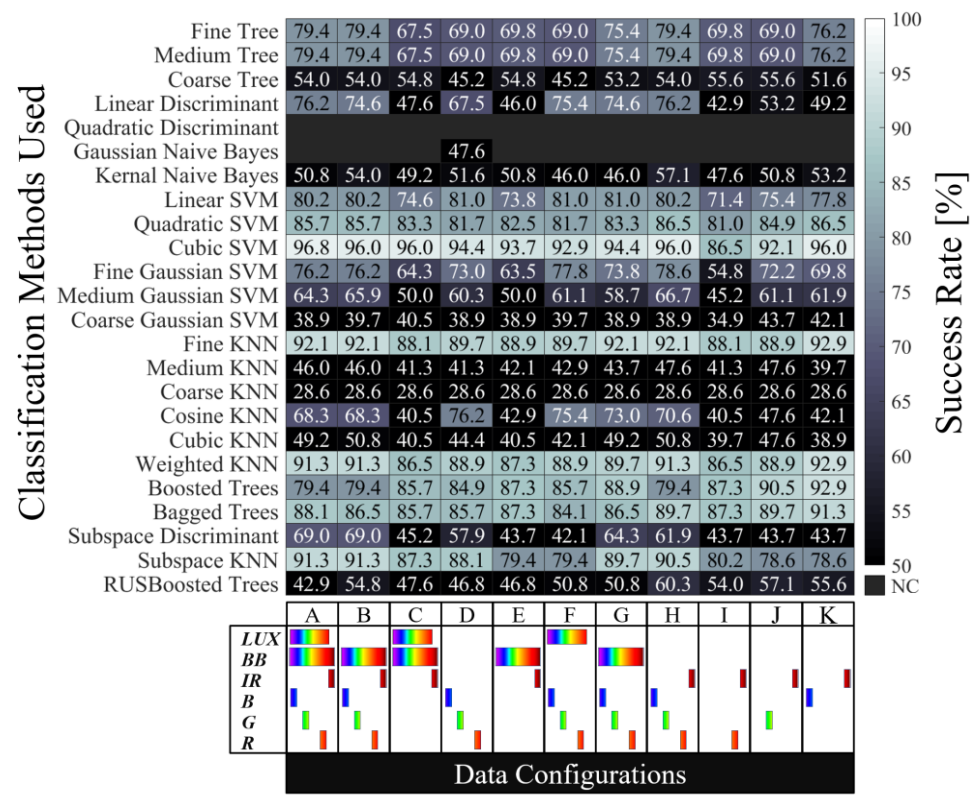

**Figure 4.** Results of training classifiers on the raw data set. The classification success varies drastically depending on the classification method used. The best results are achieved with Cubic SVM, Fine KNN and Weighted KNN. The choice of the data features configuration has a less important impact on the training results.

To understand errors produced by the trained classifiers, a 'decision surface' of each of them can be plotted. Indeed, each classification method uses a different algorithm to establish the boundaries that separate the observations of different light sources. Once trained, each classifier uses its algorithm to determine the parameters of the equations that define these boundaries. Decision surfaces help visualize the limits established by a classification method between classes. As an example, the Figure 5(a) displays the 126 data represented in a space corresponding to the *IR* versus *R* ratio. The Figure 5(b-c) illustrates the decision surface of the Cubic SVM and Fine KNN classification method in configuration 'I' (for an easy representation in 2D space). In practice, good classification success rates are more likely to be obtained when values of the classes are distinctively separated from each other. But in our case, we can see some very narrow zones with experimental dots right at the boundaries inducing errors of classification. It explains the numbers from Figure 4, that Cubic SVM Configuration I has a rate of success of 86.5% while the Fine KNN Configuration I has a rate of success of 88.1%. More decision surfaces from different classification methods can be found in Figure S1 in the supporting information. Let us note that for the configuration with more than 3 data, the representation of surface decision would have been more complex.

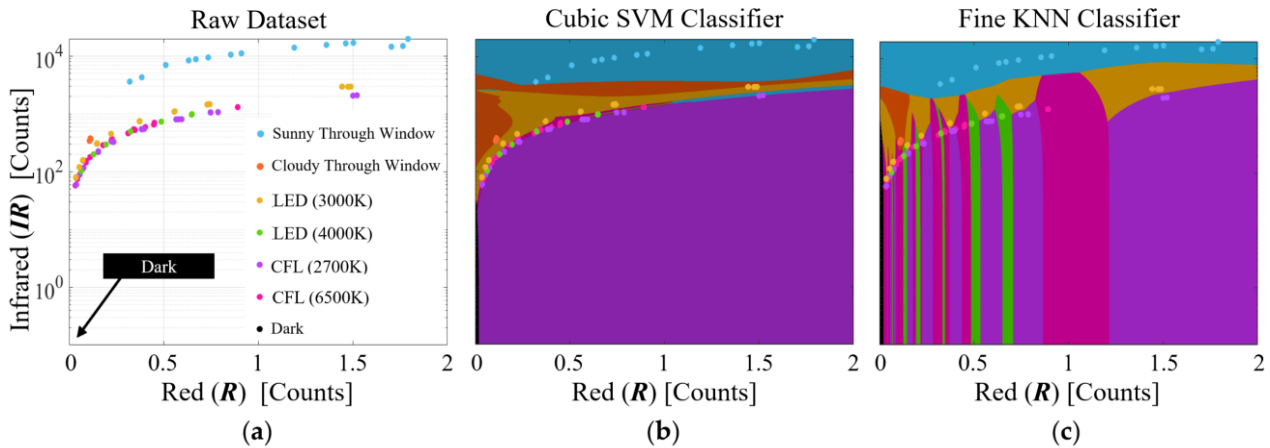

**Figure 5:** Representation of the 126 observations in the raw dataset. **(a)** Each observation is positioned according to the *R* and *IR* values. Decision surface of **(b)** the Cubic SVM classification method and **(c)** the Fine KNN classification method, both using I data configuration. Decision surfaces of the four best classifiers are shown in the supplementary Figure S1.

*3.2. Normalization Impact on Classification Performances*

Because the light source classes are not concentrated in zones distinct from each other but distributed in intensity over all their features, it is complicated for classifiers to achieve successful light source classification. The poor results obtained during training suggest an underfitting for the dataset used. Although underfitting can be attributed to the small sample size of our dataset, unlike natural light sources, artificial light emissions tend to remain constant once turned on. Consequently, a high number of samples for artificial light should not be necessary.

However, for natural light sources, the problem of the number of observations necessary to cover their intensity range remains. One way to get around this problem is to eliminate the intensity variation for the classifiers training. Applying a difference normalization to the data, it is possible to constrain the fluctuation of the data to be related to spectral variations. This method of normalization is commonly used in fields of remote sensing using multispectral or hyperspectral data. For instance, it can help to distinguish vegetation health with the Normalization Difference Vegetal Index (NDVI) [18]. This method is based on a simple calculation between two values: dividing the difference of the two values by their sum, as shown in Equation (1).

$$Normalized\ Difference\ Data = \frac{Data\ to\ Normalize - Normative\ Data}{(Data\ to\ Normalize + Normative\ Data)}. \quad (1)$$

As an example, Figure 6(a) displays the 126 data normalized to the Blue (*B*) value. Applying this method, to each measurement, a clearer map of the light source classes emerges in the Figure 6(b) which is the decision surface created using the Blue normalized dataset using the Fine KNN in configuration I classification method. With this figure we can see significant improvement in the way classes can be recognized. The decision surfaces are better defined and should allow better classification results. As a matter of fact, the classification training with the Fine KNN configuration I algorithm reaches a perfect score of 100%.

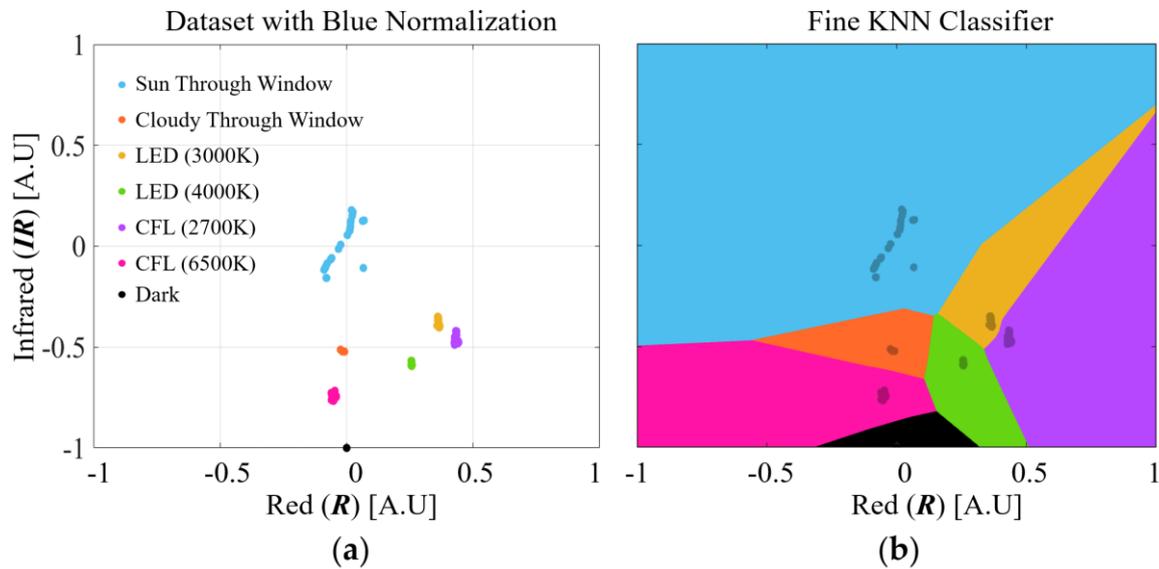

**Figure 6.** Results of the classifiers training after using difference normalization **(a)** Representation of the 126 observations using normalization to blue value. Light source observations are very distinctly separated **(b)** Applying the Fine KNN method, its decision surface exhibits a significant improvement with very clear zones for each light source class.

Tests have been conducted for all the 264 configurations to establish the impact of the Blue (*B*) normalization of the differences. As visible in Figure 6(b) for a Fine KNN classifier with Blue normalization, classifiers trained on normalized datasets are fed with classes distinctively grouped and no longer spread out according to their intensity. More decision surfaces from different classification algorithms trained for different normalization values can be found in the supplementary Figure S2. Figure 7 summarizes the scores for all the 24 classification methods with configuration from A to K configurations we have tested with Blue normalization. We can see a clear improvement of the success rates in almost all the configurations. We can also see that the K configuration gives no result. Its input is reduced to only one data, *IR*, leading to poor performances.

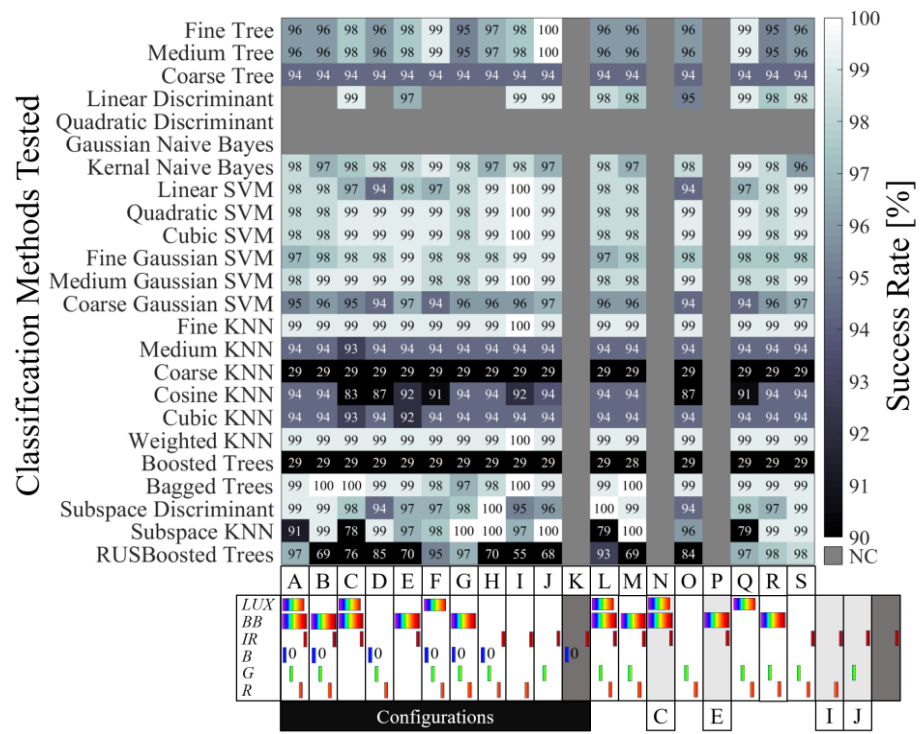

**Figure 7.** Performance of different classifiers trained through a 5-fold cross validation process with the Blue normalization.

One detail about normalizing by the difference is that if the normalization is also applied to the value chosen as the norm, it will be normalized by itself and inevitably equal to zero. Since we wanted to test the effect of these 0 values for the classification, we have tested, for each classification method, some additional configurations which are almost the same 11 initial configurations (A to H) but removing the data used for normalization. Details on the configurations for each normalization are available in the supporting Figure S3. Since it gives duplicates or relying on only one 'feature'. Finally, only 8 configurations are added (from L to S). As an example, we show in Figure 5 different configurations used for the Blue normalization and the results provided from the 5-fold cross validation. Three configurations are not tested out of the 19 because, in this case, the I configuration would have had only *IR* values, and in the case of N and P configurations, it would have reproduced the C and E configurations respectively. In total, for the Blue normalization, 384 classifiers have been tested, many of them reaching performances over 90%, and 40 of them achieved 100% success rate, mainly with the Fine and Weighted KNN methods.

Similar tests have been done with the other 'colored' normalizations. As it is shown in the complementary, the number of classifiers to be tested is the same for the Green and Red normalization (384), and a bit lower for the *BB* and *IR*, respectively 336 and 264. A total of 1752 classifiers are trained. Results of their 5-fold cross validation performances are shown in the supplementary. Over these 1752 classifiers, 267 achieved 100% correct classification which is a significant improvement compared to classification performances without normalization. Figure 8 summarizes in one table the number of classifications with 100% success, depending on the classification method applied and the feature configuration used, all normalization combined. It appears from this figure that, in this training phase, Fine KNN and Weighted KNN are the most performant classification methods, as well as the L configuration seems the most appropriate feature configuration.

| Classification Methods Used | Total | A | B | C | D | E | F | G | H | I | J | K | L | M | N | O | P | Q | R | S |
|---|---|---|---|---|---|---|---|---|---|---|---|---|---|---|---|---|---|---|---|---|
| Fine Tree | 4 | 0 | 0 | 1 | 0 | 1 | 0 | 0 | 0 | 0 | 1 | 1 | 0 | 0 | 0 | 0 | 0 | 0 | 0 | 0 |
| Medium Tree | 4 | 0 | 0 | 1 | 0 | 1 | 0 | 0 | 0 | 0 | 1 | 1 | 0 | 0 | 0 | 0 | 0 | 0 | 0 | 0 |
| Coarse Tree | 0 | 0 | 0 | 0 | 0 | 0 | 0 | 0 | 0 | 0 | 0 | 0 | 0 | 0 | 0 | 0 | 0 | 0 | 0 | 0 |
| Linear Discriminant | 18 | 0 | 0 | 3 | 0 | 2 | 2 | 0 | 1 | 0 | 1 | 1 | 5 | 3 | 0 | 0 | 0 | 0 | 0 | 0 |
| Quadratic Discriminant | 0 | 0 | 0 | 0 | 0 | 0 | 0 | 0 | 0 | 0 | 0 | 0 | 0 | 0 | 0 | 0 | 0 | 0 | 0 | 0 |
| Gaussian Naive Bayes | 0 | 0 | 0 | 0 | 0 | 0 | 0 | 0 | 0 | 0 | 0 | 0 | 0 | 0 | 0 | 0 | 0 | 0 | 0 | 0 |
| Kernel Naive Bayes | 3 | 1 | 0 | 1 | 0 | 0 | 0 | 0 | 0 | 0 | 0 | 1 | 0 | 0 | 0 | 0 | 0 | 0 | 0 | 0 |
| Linear SVM | 9 | 1 | 1 | 0 | 1 | 0 | 0 | 0 | 2 | 1 | 0 | 2 | 1 | 0 | 0 | 0 | 0 | 0 | 0 | 0 |
| Quadratic SVM | 21 | 2 | 2 | 2 | 1 | 2 | 2 | 1 | 2 | 2 | 1 | 2 | 2 | 0 | 0 | 0 | 0 | 0 | 0 | 0 |
| Cubic SVM | 23 | 2 | 2 | 3 | 1 | 2 | 2 | 1 | 2 | 2 | 1 | 2 | 2 | 0 | 1 | 0 | 0 | 0 | 0 | 0 |
| Fine Gaussian SVM | 10 | 0 | 1 | 0 | 1 | 2 | 0 | 1 | 1 | 2 | 1 | 1 | 0 | 0 | 0 | 0 | 0 | 0 | 0 | 0 |
| Medium Gaussian SVM | 22 | 2 | 2 | 1 | 2 | 3 | 1 | 2 | 2 | 2 | 1 | 2 | 2 | 0 | 0 | 0 | 0 | 0 | 0 | 0 |
| Coarse Gaussian SVM | 2 | 0 | 0 | 0 | 0 | 1 | 0 | 0 | 0 | 0 | 0 | 1 | 0 | 0 | 0 | 0 | 0 | 0 | 0 | 0 |
| Fine KNN | 39 | 5 | 5 | 2 | 2 | 3 | 2 | 5 | 2 | 2 | 1 | 2 | 5 | 3 | 0 | 0 | 0 | 0 | 3 | 0 |
| Medium KNN | 0 | 0 | 0 | 0 | 0 | 0 | 0 | 0 | 0 | 0 | 0 | 0 | 0 | 0 | 0 | 0 | 0 | 0 | 0 | 0 |
| Coarse KNN | 0 | 0 | 0 | 0 | 0 | 0 | 0 | 0 | 0 | 0 | 0 | 0 | 0 | 0 | 0 | 0 | 0 | 0 | 0 | 0 |
| Cosine KNN | 0 | 0 | 0 | 0 | 0 | 0 | 0 | 0 | 0 | 0 | 0 | 0 | 0 | 0 | 0 | 0 | 0 | 0 | 0 | 0 |
| Cubic KNN | 0 | 0 | 0 | 0 | 0 | 0 | 0 | 0 | 0 | 0 | 0 | 0 | 0 | 0 | 0 | 0 | 0 | 0 | 0 | 0 |
| Weighted KNN | 39 | 5 | 5 | 2 | 2 | 3 | 2 | 5 | 2 | 2 | 1 | 2 | 5 | 3 | 0 | 0 | 0 | 0 | 3 | 0 |
| Boosted Trees | 0 | 0 | 0 | 0 | 0 | 0 | 0 | 0 | 0 | 0 | 0 | 0 | 0 | 0 | 0 | 0 | 0 | 0 | 0 | 0 |
| Bagged Trees | 21 | 2 | 1 | 4 | 1 | 2 | 1 | 1 | 1 | 2 | 1 | 3 | 1 | 1 | 0 | 0 | 0 | 0 | 0 | 0 |
| Subspace Discriminant | 20 | 4 | 2 | 2 | 0 | 0 | 1 | 1 | 2 | 0 | 0 | 1 | 5 | 2 | 0 | 0 | 0 | 0 | 2 | 0 |
| Subspace KNN | 19 | 0 | 5 | 0 | 2 | 0 | 0 | 5 | 4 | 0 | 1 | 0 | 0 | 2 | 0 | 0 | 0 | 0 | 3 | 2 |
| RUSBoosted Trees | 0 | 0 | 0 | 0 | 0 | 0 | 0 | 0 | 0 | 0 | 0 | 0 | 0 | 0 | 0 | 0 | 0 | 0 | 0 | 0 |
| Total | | 24 | 26 | 22 | 13 | 22 | 13 | 22 | 21 | 15 | 11 | 21 | 29 | 14 | 1 | 0 | 0 | 0 | 11 | 2 |

Data Configurations

**Figure 8.** Summary of classifiers and configurations reaching 100% correct classification, all normalizations combined.

*3.3. Successfully Trained Classifiers in Confrontation to New Controlled Pseudo-Spectra*

In the next step, the 256 best classifiers, have been confronted with new observations from the compact low-cost device. These observations were made in a controlled light environment consisting of several single light sources, for which classifiers have been trained: LED 3000K, LED 4000K, CFL 2700K, CFL 6500K and cloudy sky behind a window at different light intensities, as seen in Figure 9(b). After analysis, the accumulated data from the compact device are provided to the classifiers to obtain their classification results. Even if trained in similar conditions, it is not obvious all classifiers will be successful, mainly because of the overfitting effect. This means that the classifier is overfitted to the training data it was given and is not able to correctly recognize data that is not part of that data. This classifier is therefore not able to generalize its classification method to other data that are similar but not identical to the data used for its training. The results obtained on the new controlled environment are shown in Figure 9(a). Let us note that the column from the configuration O, P and Q have been removed since no classifier was able to reach the score of 100% after the 5-fold cross validation process. For the same reason, we have removed some lines which represent unreliable classification methods like 'coarse tree', Quadratic discriminant, Gaussian Naïve Bayes, some KNN (Medium, Coarse, Cosine and Cubic), Boosted tree and finally the RUSBoosted tree.

The main result of this experiment shown on Figure 9(a) is that among the classifiers that reach 100% correct classifications in training, more than half the classifiers have 99% of success recognizing the new spectra were trained for, and 42 classifiers reach the performance of 100% of good classification, mainly from the Green and Blue normalization. Classification results for all normalization are grouped in the supporting Figure S4.

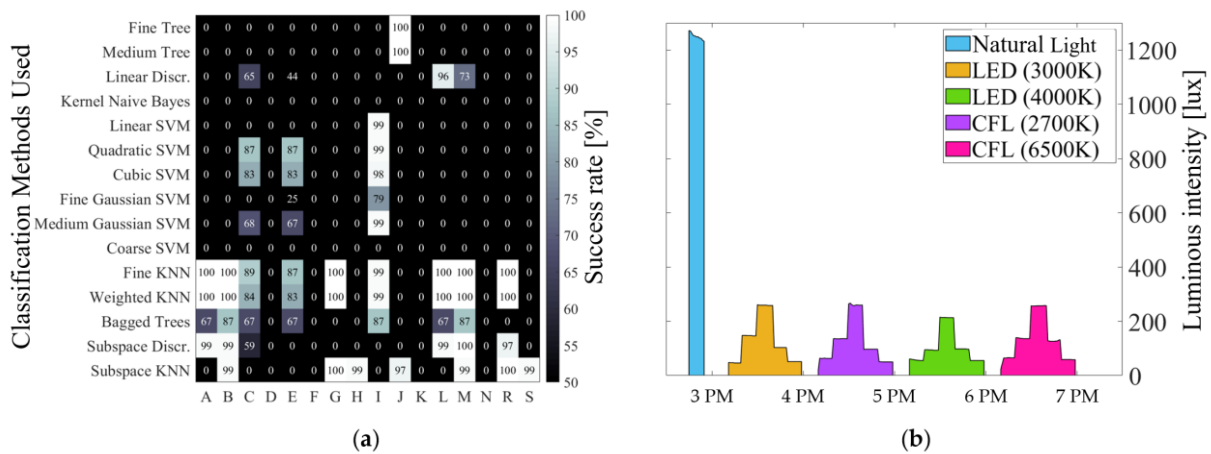

**Figure 9. (a)** Classification results using Blue normalization in a controlled light environment. **(b)** Controlled light environment used to create generalization test data for classifiers that passed the training test.

*3.4. Trained Classifiers performances on Pseudo-Spectra from Uncontrolled Light Environment*

To go further on testing the 42 classifiers that have successfully classified light sources in controlled conditions, it is necessary to evaluate their performance in real 'uncontrolled' indoor conditions. In such conditions, the light is usually a mix of artificial and natural light sources. The problematic here is that the nature of a real light environment is richer in spectrum and intensity variations in the real environment tested than the reduced environmental data used for training and generalization of classifiers. This makes the comparison of performance between each classifier more delicate than in the previous steps.

The method chosen to evaluate the performance of the classifiers used here is based on two criteria. On the one hand, a classifier is considered valid if, as in the previous steps, it correctly recognizes a light source when it is the only source present in the environment. On the other hand, when the light environment is a mix of several sources, the classifiers are evaluated on its ability to distinguish which source of radiation is the main contributor in terms of irradiance. In other words, a light source must be recognized as the main light source when it emits more than half of the irradiance measured in the environment. For example, during the day, the natural light is usually predominant, compared to the artificial light. On that criterion, a classifier would be considered performant if switching from natural light to artificial light as main light source as soon as daylight would have decreased under 50% of the total power radiated. For this experiment, we have tested an indoor environment globally composed of two types of light sources: a LED 3000K and a window facing south-east with a clear blue sky. For several days of tests, the LED 3000K, constant non-tunable lighting, was kept on during the entire working day, making it easier to observe the switching ratio.

After several days of experimentation, results show that only 23 classifiers over the 42 tested could distinguish the different classes properly observed in the real indoor environment: LED 3000K, blue sky through a window and dark. During these tests, a precise measurement of the irradiance was carried out with a compact spectrometer. This measurement allows calculating, at each moment of the day, the percentage of the total incident radiation emitted by the artificial LED 3000K and by the natural light through a window. As a result, when a classifier switches from a recognized source to another, the ratio of natural light versus artificial light is known. As a criterion we have established earlier, the more this switching point occurs near 50%, the more the classifier tested will be considered as reliable.

Figure 10 shows the classification results of a typical day, for the two best classifiers both using the "Fine KNN" method with the blue normalization using the *G* and *R* configurations. On average, over different days, these classifiers were able to switch consistently from one source to the other with switching ratios around 45%. Meaning that, if the irradiance emitted by the LED in the environment is larger than 45% of the total irradiance radian, the classifier recognizes this source as the main one. Otherwise, the classifier determines that the environment is composed of natural light through a window. The average switching ratio of all the other classifiers is shown in the supplementary Figure S5. From these results, K-nearest neighbor (KNN) methods of classification seems appropriate to our application in combination Red or Blue normalization.

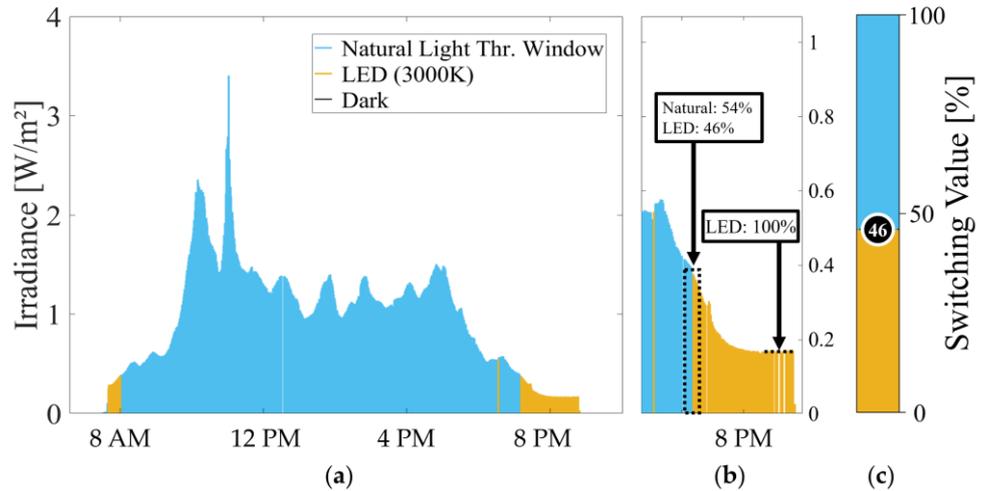

**Figure 10. (a)** Classes recognized for a day with mixed light source using Fine KNN method, associated with the R data configuration; **(b)** Switching moment between natural light and LED light classes; **(c)** Switching ratio value between those two classes.

## 4. Harvestable Energy Calculation

A proper calculation of the harvestable energy in specific locations requires pseudo-spectra to be classified correctly. Then from the class recognized, a full spectrum can be 'reconstructed' to meet the resolution of a spectrum acquired by a spectrometer. This 're-construction' process relies on different pieces of data: reference spectra acquired by a compact spectrometer and the light intensity (*LUX*) data acquired by our low-cost sensor device. First, the intensity of the spectrum must be tuned such as its illuminance (irradiance weighted according to the photopic luminous efficiency function) corresponds to the *LUX* value. Then, the harvestable energy is then estimated as described in Politi 2021. The light source classification method used in the following is one of the most efficient we've seen in the previous chapter, the Fine KNN with normalization by Blue and associated with configuration R.

### 4.1. Sources of Error for Spectra Reconstruction

As written previously, a good spectrum reconstruction needs a reliable value of light intensity. For this we use the data *LUX* from the low-cost TSL2561 sensor. But when comparing the lux value from this sensor to the reference lux calculated from an expensive calibrated spectrophotometer a difference as important as 30% can be measured, as seen in Figure 10(a). Depending on the class of light recognized, a correction factor is applied to the lux data obtained by the TSL2561. This correction aims to reduce the inconsistency of the sensor. Another way to address this problem would be to create a model of the photodiodes composing this sensor to develop a digital twin, but this topic lies out of the scope of this paper. During these experiments, we have faced an additional source of error. Indeed, while the correction factor is set to be constant whatever the light intensity for the artificial sources, for the natural light a constant correction factor cannot be applied.

Indeed, under high illuminance levels (> 1500 lux) or at low illuminance at the very beginning and the end of the day, the natural light spectra are substantially different (see Figure S6 in supplementary). These variations can induce errors in the luxmeter readings up to 100% as seen is Figure 11(b). In conclusion of these observations, natural light sources must be decomposed in multiple light source classes. Figure 11(b) shows the strong difference in the error given by the luxmeter exposed to natural light coming through a window. Four distinct light classes can be defined corresponding to the different change in the light composition throughout the day: sunrise, sunset, daylight, and strong daylight. Classifiers can be trained again to consider the new sources. To complete this improvement of the analysis system, these four types of light sources are each associated with correction functions. These functions are defined using regression method because of the non-constant and non-linear nature of the sensor error under natural light seen in Figure 11(b). Correction coefficients and functions are then integrated to supply spectrum reconstruction process with more precise lux.

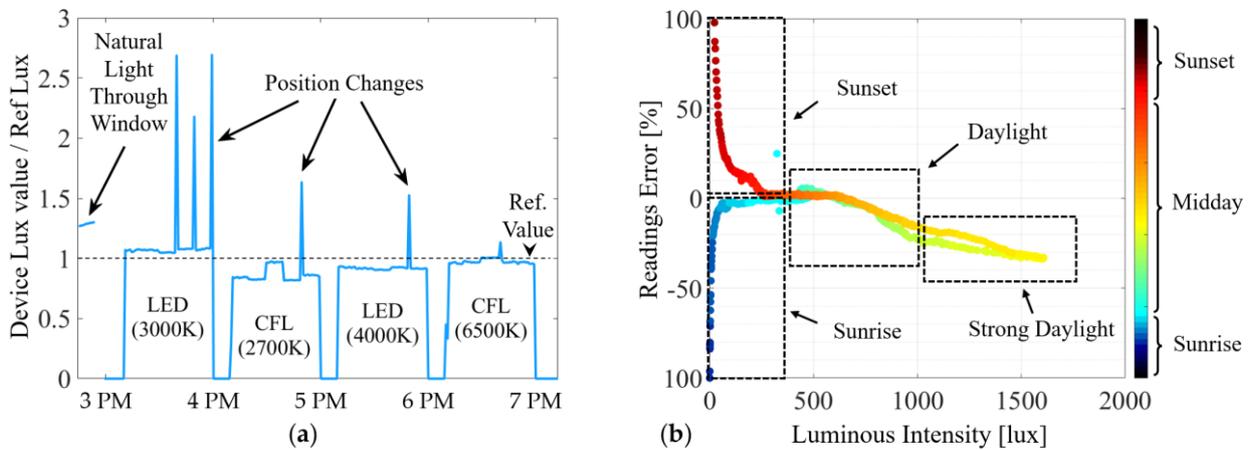

**Figure 11. (a)** Light intensity values given by the TSL2561 sensor in lux, values normalized to the lux value given by the spectrometer. Values from the sensor depend on the type of light source and its intensity and can induce errors. **(b)** Evolution of the low-cost system's light intensity measurement error as a function of the spectrometer's reference light intensity. Measurements made during a whole day in an environment mainly lit by natural light passing through a window. Selection of different natural light sources to add to the classification system for greater accuracy.

*4.2. Harvestable Energy Calculation Results Using the Low-Cost Sensor*

Finally, with our complete system outputting reconstructed spectra, the calculation model results can be studied. The low-cost system has been tested over a period of 16 consecutive days. For comparison purpose, the compact spectrometer was installed next to the prototype. Those tests' goal is to verify whether the classification and spectrum reconstruction processes of our system are reliable and consistent. Figure 12 displays the results of the model calculations based on the low-cost system as well as its calculations based on the compact spectrometer readings. As for control results, power and energy practical measurements are shown. More results for other days are gathered in the Figure S7 from the supplementary material. These measurements were conducted by the energy harvesting prototype placed near the two analysis devices. The fine KNN classification method was used in association with the feature configuration R and the Blue normalization. For a given day presented in the figure, the error between energy harvestable calculations based on the spectrum reconstruction process and the compact spectrometer is only 1.4%. Over the sixteen-day observation samples, this average percentage of absolute error on the energy harvestable is 3.4%. This confirms that the light source classification method, combined with the spectra reconstruction, can provide model calculation results with a similar level of accuracy than results described in Politi et al. [15] that use an expensive compact spectrophotometer. This is an encouraging result, and it suggests that this method can compete with a higher resolution measurement system. This level of accuracy, for a system using a relatively simple computational model and a low-cost measurement device, is an interesting support to help evaluate the sizing needed for an energy recovery device to be able to make a consumer device self-sufficient.

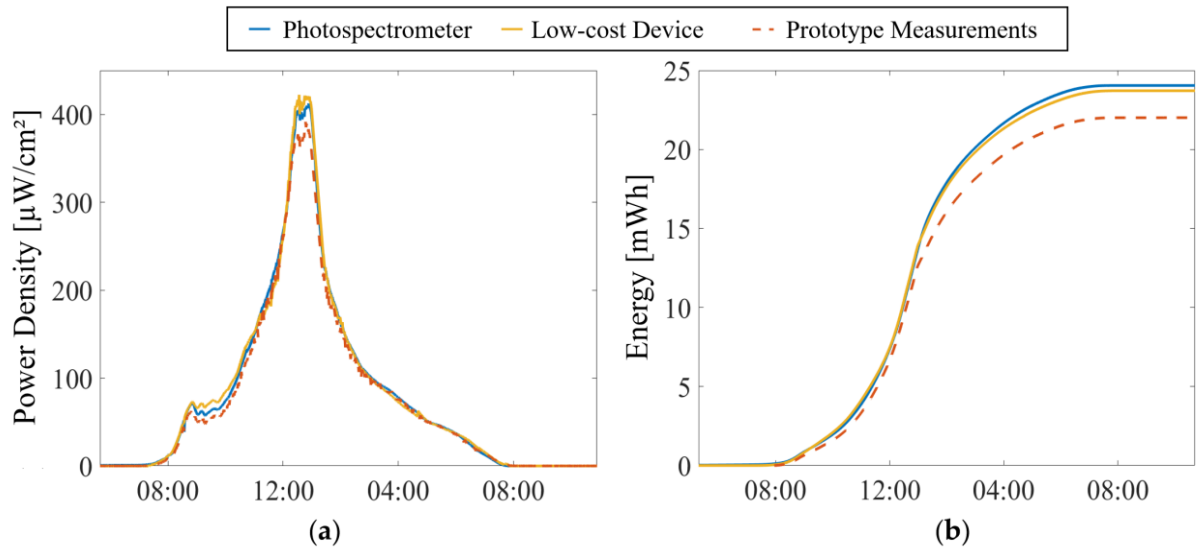

**Figure 12.** Result of the reconstruction method, after implementing a more accurate correction of the measured values by the low-cost light intensity system.

As a proof of concept, a graphical interface described in the Figure S8 shows how these results could be exploited by a user. The main results of the calculations of the recoverable energy as well as the minimum surface area and number of solar cells (GaAs solar cells in our example) to compensate for the consumption of the electric device to supply in energy are displayed. As an example, the recommendation after the 16 days of observation is to use 283 cm² of GaAs solar cells to supply in energy a wireless e-ink tablet consuming an average power per day of about 10 mW.

## 5. Conclusion

In this work, we have explored for the first time how an ultra-low-cost light sensor device capable of estimating the harvestable energy in real indoor conditions. This device coupled to supervised learning methods of classification method offers a powerful low-cost alternative to other more conventional methods of light characterization. This low-cost system can guide PV sizing with precision comparable to a system using an expensive spectrometer. The mean absolute error percentage between experimentally harvested light energy by a harvesting prototype and the model calculation based on spectrometer data or the low-cost device is lower than 6%. To go even further, tests over longer periods will be realized to ensure that this system could apply its calculations throughout the year, regardless of the variations in light due to seasonal changes. Implementing more precise/complex calculation model a larger number of different PV technology can also be added in our database to give the user the choice of the solar cell. For further improvements, the development of a digital twin of the sensor could help with correction of its readings. Finally, for the moment, our proof-of-concept works using Matlab pretreatment. By minimizing the energy required to use a classifier, it would be feasible to use it directly on our measurement system, as demonstrated in the work of Micheals et al. [19], and even to train a classifier in an embedded way.

As said, many improvements could be done, but we believe our work opens an opportunity for engineers and researchers to deploy many autonomous devices in indoor conditions by using this method.

**Supplementary Materials:** The following are available online at https://doi.org/10.5281/zenodo.5807743, **Figure S1:** Surfaces decision of the most successful trained classifiers, **Figure S2:** Decision surface obtained for different classifiers with different normalizations, **Figure S3:** Details on the values featured in the data configuration for each normalization, **Figure S4:** Generalization results of the classifiers for the different types of normalization applied to the data, **Figure S5:** Value of the switching percentage of the classifiers tested., **Figure S6:** Difference between four light spectra taken at different time of the day, **Figure S7:** Comparison between the energy harvestable calculated based on the spectrometer data and the low-cost device with the prototype measurement as control, **Figure 8:** Example of the results obtained using an graphical user interface applying the method exposed in the article.

**Author Contributions:** Conceptualization, B.P. and N.C.; methodology, B.P.; software, B.P.; validation, B.P. and N.C.; formal analysis, B.P.; data curation, B.P.; writing—original draft preparation, B.P.; writing—review and editing, B.P. and N.C.; supervision, A.F. and N.C.; project administration N.C. and A.F.; funding acquisition, N.C. and A.F. All authors have read and agreed to the published version of the manuscript.

**Funding:** This research was funded by the Association Nationale de la Recherche et de la Technologie (ANRT), grant number CIFRE 2017/0331 and by the company Bureaux A Partager. The APC was funded by the EPF Engineering School of Montpellier.

**Conflicts of Interest:** The authors declare no conflict of interest.

**References**

1. Carvalho, C.; Paulino, N. On the Feasibility of Indoor Light Energy Harvesting for Wireless Sensor Networks. Procedia Technol. 2014, 17, 343–350, doi:10.1016/j.protcy.2014.10.206.
2. Wang, Y.; Liu, Y.; Wang, C.; Li, Z.; Sheng, X.; Lee, H.G.; Chang, N.; Yang, H. Storage-less and converter-less photovoltaic energy harvesting with maximum power point tracking for internet of things. IEEE Trans. Comput. Des. Integr. Circuits Syst. 2015, 35, 173–186.
3. European Commission- Joint Research Centre (JRC) Photovoltaic geographical information system (PVGIS) Available online: https://ec.europa.eu/jrc/en/pvgis.
4. Minnaert, B.; Veelaert, P. A proposal for typical artificial light sources for the characterization of indoor photovoltaic applications. Energies 2014, 7, 1500–1516, doi:10.3390/en7031500.
5. Li, Y.; Grabham, N.J.; Beeby, S.P.; Tudor, M.J. The effect of the type of illumination on the energy harvesting performance of solar cells. Sol. Energy 2015, 111, 21–29.
6. Randall, J.F. On the use of photovoltaic ambient energy sources for powering indoor electronic devices, 2003.
7. Bader, S.; Ma, X.; Oelmann, B. One-diode photovoltaic model parameters at indoor illumination levels – A comparison. Sol. Energy 2019, 180, 707–716, doi:10.1016/j.solener.2019.01.048.
8. Randall, J.F.; Jacot, J. Is AM1.5 applicable in practice? Modelling eight photovoltaic materials with respect to light intensity and two spectra. Renew. Energy 2003, 28, 1851–1864, doi:10.1016/S0960-1481(03)00068-5.
9. Hande, A.; Polk, T.; Walker, W.; Bhatia, D. Indoor solar energy harvesting for sensor network router nodes. Microprocess. Microsyst. 2007, 31, 420–432, doi:10.1016/j.micpro.2007.02.006.
10. Afsar, Y.; Sarik, J.; Gorlatova, M.; Zussman, G.; Kymissis, I. Evaluating photovoltaic performance indoors. Conf. Rec. IEEE Photovolt. Spec. Conf. 2012, 1948–1951, doi:10.1109/PVSC.2012.6317977.
11. Teran, A.S.; Wong, J.; Lim, W.; Kim, G.; Lee, Y.; Blaauw, D.; Phillips, J.D. AlGaAs Photovoltaics for Indoor Energy Harvesting in mm-Scale Wireless Sensor Nodes. IEEE Trans. Electron Devices 2015, 62, 2170–2175, doi:10.1109/TED.2015.2434336.
12. Mathews, I.; King, P.J.; Stafford, F.; Frizzell, R. Performance of III – V Solar Cells as Indoor Light Energy Harvesters. IEEE J. Photovoltaics 2016, 6, 230–235, doi:10.1109/JPHOTOV.2015.2487825.
13. Freitag, M.; Teuscher, J.; Saygili, Y.; Zhang, X.; Giordano, F.; Liska, P.; Hua, J.; Zakeeruddin, S.M.; Moser, J.E.; Grätzel, M.; et al. Dye-sensitized solar cells for efficient power generation under ambient lighting. Nat. Photonics 2017, 11, 372–378, doi:10.1038/nphoton.2017.60.
14. Müller, M.; Wienold, J.; Walker, W.D.; Reindl, L.M. Characterization of indoor photovoltaic devices and light. Conf. Rec. IEEE Photovolt. Spec. Conf. 2009, 000738–000743, doi:10.1109/PVSC.2009.5411178.
15. Politi, B.; Parola, S.; Gademer, A.; Pegart, D.; Piquemil, M.; Foucaran, A.; Camara, N. Practical PV energy harvesting under real indoor lighting conditions. Sol. Energy 2021, 224, 3–9, doi:10.1016/j.solener.2021.05.084.
16. Sarik, J.; Kim, K.; Gorlatova, M.; Kymissis, I.; Zussman, G. More than meets the eye - A portable measurement unit for characterizing light energy availability. 2013 IEEE Glob. Conf. Signal Inf. Process. Glob. 2013 - Proc. 2013, 387–390, doi:10.1109/GlobalSIP.2013.6736896.

17. Ma, X.; Bader, S.; Oelmann, B. Characterization of indoor light conditions by light source classification. IEEE Sens. J. 2017, 17, 3884–3891, doi:10.1109/JSEN.2017.2699330.
18. Beck, P.S.A.; Atzberger, C.; Høgda, K.A.; Johansen, B.; Skidmore, A.K. Improved monitoring of vegetation dynamics at very high latitudes: A new method using MODIS NDVI. Remote Sens. Environ. 2006, 100, 321–334, doi:10.1016/j.rse.2005.10.021.
19. Michaels, H.; Rinderle, M.; Freitag, R.; Benesperi, I.; Edvinsson, T.; Socher, R.; Gagliardi, A.; Freitag, M. Dye-sensitized solar cells under Ambient Light Powering Machine Learning: Towards autonomous smart sensors for Internet of Things. Chem. Sci. 2020, 11, doi:10.1039/c9sc06145b.